# *Ab initio* study of noncentrosymmetric transition-metal monogermanide B20-RhGe synthesized at high temperature and pressure

N.M. Chtchelkatchev[1,2,a], M.V. Magnitskaya[1,3,b], and A.V. Tsvyashchenko[1]

[1] Vereshchagin Institute for High Pressure Physics, Russian Academy of Sciences, 108840 Troitsk, Moscow, Russia
[2] Moscow Institute of Physics and Technology, 141700 Dolgoprudny, Moscow Region, Russia
[3] Lebedev Physical Institute, Russian Academy of Sciences, 119991 Moscow, Russia

**Abstract.** We present *ab initio* density-functional study of the noncentrosymmetric B20-type phase of RhGe, which is not found in nature and can be synthesized only at extreme pressures and temperatures. The structural, thermodynamic, electronic, lattice-dynamical, and transport properties of B20-RhGe are calculated, and their evolution with increasing pressure is traced. The temperature dependence of the charge and heat transport properties is evaluated within the semi-classical Boltzmann approach. Using the quasi-harmonic approximation, we determine the range of pressures and temperatures, in which B20-RhGe is stable, and make recommendations for optimizing the synthesis conditions in order to reduce the number of defects that occur in a sample during solidification.

## 1 Introduction

Creation of new materials is always topical, as it opens up prospects for technological development in various fields. Therefore, the method of manufacture and processing is of particular importance, since it determines the structure of materials and, hence, their properties and possible applications. One of the effective ways for creating novel compounds is the direct synthesis by crystallization from molten constituents at high pressures and temperatures close to those existing in the Earth's mantle. This method has been successfully applied to prepare previously unknown compounds that remain metastable at normal conditions (see, e.g., [1, 2]).

As an example, we mention transition metal (TM) monogermanides with noncentrosymmetric B20-type crystal structure, which are now actively studied [3–9], as well as isostructural TM monosilicides. The lack of central symmetry results in a wide spectrum of unusual properties. For instance, these systems include room-temperature thermoelectrics [9–11] and chiral magnets interesting to spintronics [3–5, 8, 12]. TM monosilicides are better understood than monogermanides, as they are formed at normal conditions and it is much easier to grow and study them.

As is known, high-pressure experiments are very laborious and time-consuming. Further, various defects and residual microstresses typical of nonequilibrium metastable phases can lead to uncertainties in the experimental results. In such a situation, numerical *ab initio* simulations successfully complement the experiment.

In this paper, we study structural, thermodynamic, electronic, lattice-dynamical, and transport properties of B20-type RhGe, including their evolution with pressure and temperature. Three crystalline phases of RhGe are known: one normal-pressure orthorhombic phase B31 (MnP type) and two high-pressure cubic phases B20 (FeSi type) and B2 (CsCl type). B31-RhGe crystallizes congruently from a melt at normal pressure, its melting temperature 1550 K [13]. The B20 phase exists above 7.7 GPa and the B2 phase is stable at much higher pressures [12].

The B20 phase of RhGe has been synthesized (see [12] and references therein) at $P = 8$ GPa from stoichiometric mixture of the components in the Toroid high-pressure device [14]. The pellets of well-mixed powdered constituents were placed in rock-salt pipe ampoule. The sample was directly heated electrically to above the melting temperature of the mixture (1700 K), the melting temperature controlled by using the voltage–current plot. Then the sample was rapidly quenched to room temperature and, after solidification, the pressure was released. Unlike some other TM monogermanides, which exhibit dendritic growth [15], RhGe grows equiaxially. The obtained samples are polycrystalline and remain metastable at normal pressure for a long time.

---

[a] e-mail: n.chtchelkatchev@gmail.com
[b] e-mail: magnma@gmail.com

## 2 Calculation procedure

Our *ab initio* computations are based on the density functional theory (DFT). We used the projected-augmented-wave (PAW) pseudopotential method as implemented in the Quantum Espresso package [16], with the PBEsol version [17] of the generalized-gradient approximation (GGA) for the exchange-correlation potential. We have chosen the plane-wave kinetic cut-off energy of 100 Ry and a uniform grid of 24×24×24 **k**-points for sampling the Brillouin zone. The geometry relaxation has been continued, until the residual atomic forces have been converged down to 5 meV/Å. The total energy convergence was better than $10^{-7}$ Ry/cell. The lattice-dynamics calculations have been made using the interatomic force constants approach. The thermodynamic quantities have been determined in the quasi-harmonic approximation (QHA) using the PHONOPY code [18]. Further on, we have evaluated the charge and heat transport properties as functions of temperature within the semi-classical Boltzmann approach in the relaxation-time approximation. For this purpose, the BoltzTraP2 (Boltzmann Transport Properties) package [19] has been employed, which utilizes the electronic-structure related quantities obtained with the Quantum Espresso package.

## 3 Results

### 3.1 Crystal structure and the phase stability

First, we consider the structural properties and thermodynamic stability of B20 RhGe. The unit cell of the simple cubic B20 structure is shown in Fig. 1, together with the Brillouin zone (BZ). The B20 lattice can be considered as a pairing-type distortion of an underlying rock-salt B1 (NaCl-type) structure. The TM and metalloid atom coordinates in the undistorted B1 structure are 1/4 and 3/4, while in the so called 'ideal' B20 lattice (see, e.g., [20, 12]), they are $1/4\tau = 0.1545$ and $1-1/4\tau = 0.8455$, where $\tau = (1+\sqrt{5})/2$ is the golden mean. The distortion from the ideal to the real B20 structure corresponds to a dimerization of the bonds along the <111> direction, with alternating elongation and reduction of distances, which breaks the inversion symmetry. This leads to a stronger hybridization between TM-d and Ge-p states, thereby enhancing the covalent character of the TM–Ge bonds and hence the cohesive energy. Thus, the reason for stability of the complex low-symmetry B20 structure over the high-symmetry B1 and B2 structures is the enhanced covalent character of its interatomic bonding [21].

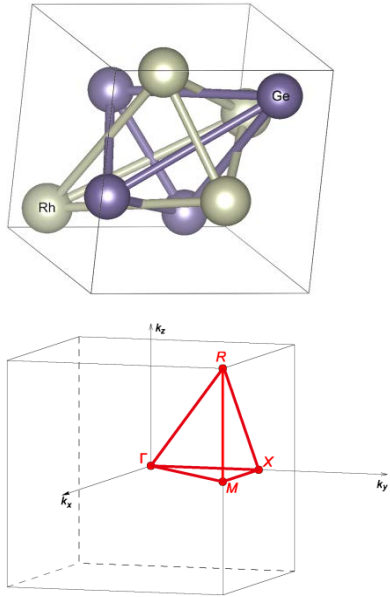
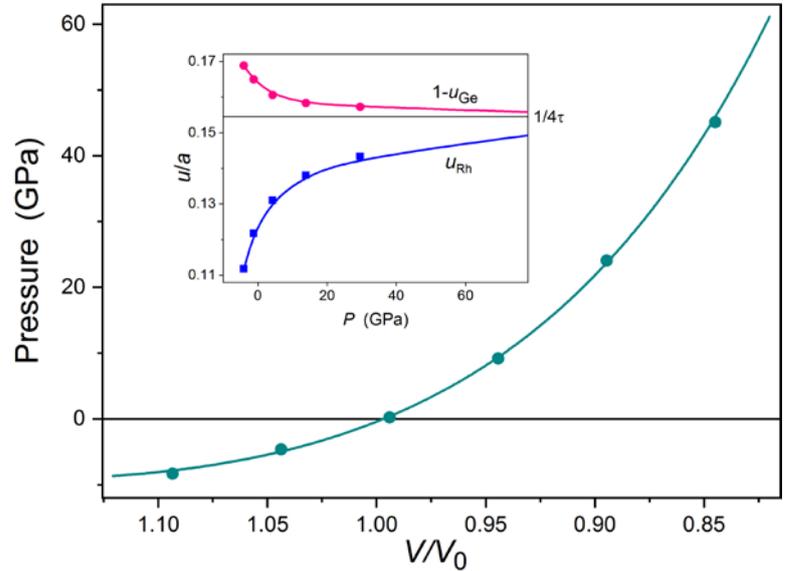

**Fig. 1.** The unit cell (top) and the Brillouin zone (bottom) of the simple cubic B20 structure (space group P2$_1$3).

**Fig. 2.** The $T = 0$ equation of state of B20-RhGe built using the Birch–Murnaghan fit ($V_0$ is the experimental volume at 0.1 MPa). Inset: atomic coordinates as functions of pressure; the horizontal line indicates their asymptotic value $1/4\tau$ in the 'ideal' B20 structure.

The X-ray diffraction at normal conditions gives for B20-RhGe the lattice constant $a = 4.8595$ Å and the atomic coordinates $u_{Rh} = 0.1281$ and $u_{Ge} = 0.8337$ [12]. Our calculated lattice parameter $a = 4.8514$ Å, which gives the equilibrium specific volume only by 0.5% smaller than the measured value. The atomic positions $u_{Rh} = 0.1321$ and $u_{Ge} = 0.8387$ are also in good agreement with experiment.

Figure 2 shows the calculated zero-temperature equation of state, $P(V/V_0)$, where $V_0$ is the experimental equilibrium volume at $P = 0.1$ MPa. According to the Birch–Murnaghan fit, the bulk modulus and its pressure derivative are $B = 202$ GPa and $B' = 9.5$, respectively. There is no available information to compare this result to. Inset to Fig. 2 schematically shows the pressure dependence of atomic coordinates ($u_{Rh}$ and $1-u_{Ge}$), which at very high pressures tend asymptotically towards their ideal-B20 value of 0.1545 indicated by the horizontal line.

Our total-energy calculations confirm that the normal-pressure ground state of RhGe is the orthorhombic B31 structure. The B20 phase becomes energetically favorable at a pressure of 7.5 GPa. Up to 100 GPa, the B20 phase remains a high-pressure ground state. The B31–B20 transition is accompanied by a decrease in the specific volume of about 5.4%, which is close to the experimental value of 4.4% [12]. On the whole, our results on the structure and stability of RhGe at $T = 0$ K are consistent with available information [5, 12, 22].

We have also evaluated relative stability of the B31-RhGe and B20-RhGe phases at nonzero temperatures in the QHA with the PHONOPY code [18]. Our preliminary results demonstrate that in the $P$–$T$ phase diagram, the B31–B20 boundary looks almost linear. It starts from 7.5 GPa at $T = 0$ K and shifts towards lower pressures with increasing temperature. Thus, the higher the temperature is, the lower the transition pressure becomes. It is known that after annealing at 1000 K, the B20 phase is transformed to the normal-pressure B31 phase. According to the calculated phase boundary, this temperature corresponds to a pressure of about 3 GPa. This suggests that the B20 phase could be synthesized at lower pressures of 4–5 GPa, which could reduce the structural inhomogeneities depending on the process of solidification under pressure. This is consistent with the synthesis conditions used in paper [23]. Hopefully, this reduction in pressure will enable to synthesize B20-RhGe single crystals. Such technique has been already developed for MnGe [7] and FeGe [8]. Our detailed results on the $P$–$T$ phase diagram of RhGe will be published elsewhere.

### 3.2 Electronic properties

Figure 3 (left) depicts the electron density of states (DOS), $N(E)$, at relative volumes $V/V_0 = 1.0$, 0.95, 0.9, and 0.85 ($P = 0.1$ MPa, 8 GPa, 22 GPa, and 43 GPa, correspondingly). In the energy range of interest ($E_F \pm 5$ eV), the DOS is contributed mostly by the hybridized Rh-4d and Ge-3p states, with domination of the former. As is seen in Fig. 3 (right), the DOS at the Fermi level, $N(E_F)$, behaves nonmonotonically with increasing pressure. The DOS shape is similar to those previously calculated for other B20-type monosilicides and monogermanides, which implies that the rigid band approximation (RBA) is valid for this class of compounds. The vertical dashed lines in Fig. 3 indicate the position of $E_F$ in B20-RuGe and hypothetical (unstable) B20-PdGe, which have,

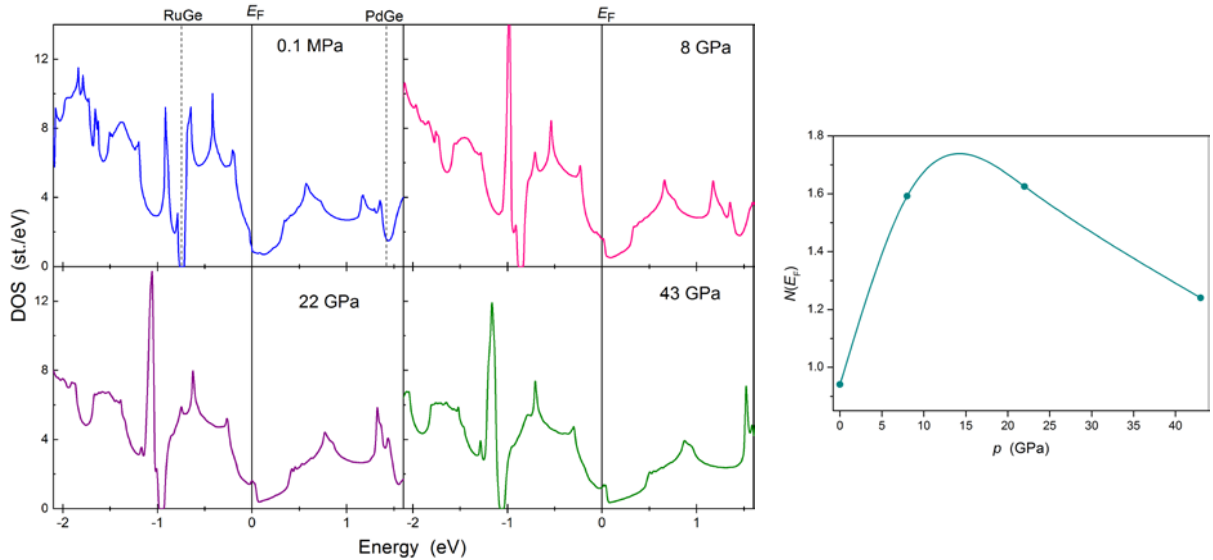

**Fig. 3.** Left: Density of states, $N(E)$, of B20-RhGe at $P = 0$, 8, 22, and 43 GPa. The Fermi energy, $E_F$, is set to zero and shown by the vertical solid line. The dashed lines indicate the $E_F$ of RuGe and PdGe. Right: pressure evolution of the $N(E_F)$.

respectively, less and more by one valence electron per formula unit.

A defining feature of the RhGe DOS is a pseudogap near the Fermi level: the $E_F$ lies in a wide valley of the DOS, and a rather low $N(E_F)$ implies that RhGe is a poor metal. In the case of semiconducting RuGe, the Fermi level falls on a 4d–3p hybridization gap. In RhGe this gap, which is due to the splitting of the d band by a low-symmetry crystal field, shifts towards lower energies with increasing pressure. In a sense, the (relatively low) pressure acts as a low-level hole doping (with Ru, Fe, Mn, etc.).

The band structure of cubic RhGe is presented in Fig. 4 (left). The bands computed at other pressures (8, 22, and 43 GPa) look very similar. The shape of the B20-type band structure is due to the nonsymmorphic and noncentrosymmetric nature of the $P2_13$ space group. This specific symmetry leads to an uncommon triply degenerate state at the zone center Γ-point. One of three degenerate bands has a rather low dispersion and zero velocity $\partial E(\mathbf{k})/\partial \mathbf{k} = 0$, while two other bands have $k$-linear dispersions and non-zero velocities equal in absolute value and opposite in sign (similar to the Dirac cone in graphene). According to the RBA, the triply degenerate state is located just below $E_F$ in the B20 compounds isovalent to RhGe and above $E_F$ in compounds with less electron count. An asymmetric energy dependence of the DOS near $E_F$ is due to the presence of this triply degenerate state.

Further, we have explored evolution of band structure and Fermi surface with increasing pressure. A change in the position of Dirac-point feature on compression is illustrated in Fig. 4 (right), where the relevant area of the band diagram along the **k**-path M–Γ–R is shown on an enlarged scale. The Dirac-point feature is seen to cross the Fermi level at 22 GPa.

Figure 5 illustrates the pressure evolution of the Fermi surface (FS). At $P = 0.1$ MPa, the FS consists of the Γ-centered electron pocket associated with the Dirac-point feature, electron pockets at the BZ corner R, and hole pockets centered at the M-point. Very small hole pockets are also discerned in the ΓR direction. All the pockets are rather small and the FS resembles that of a semimetal. At 8 GPa, a Γ-centered cage-like structure of holes appears. It is associated with the flat band (one of the three degenerate bands) that crosses the Fermi level in the vicinity of Γ-point. As a whole, the Fermi surface of RhGe at 8 GPa is very much like the ambient-pressure FS of CoGe, which already includes the cage-like sheet [6]. This is apparently a consequence of larger specific volume of RhGe. Upon further compression, the cage-like structure develops mainly in the Γ–R and Γ–M directions, and at 43 GPa eventually becomes a closed simply-connected surface.

Thus, the compression to 22 GPa causes void-type Lifshitz electronic topological transitions (ETP), when electron and hole sheets of FS appear and disappear. Also noteworthy is a neck-type ETP between 22 and 43 GPa, when a change in the FS connectivity takes place. Such transitions can be observed, for instance, in quantum oscillations (see, e.g. [24]). Another way of observing the ETPs — measurements of electrical resistivity and thermopower — is discussed in Sec. 3.4.

## 3.3 Lattice dynamics

Our enthalpy calculations (Sec. 3.1) confirm the thermodynamic stability of the B20-RhGe against the polymorphic transitions to other crystal phases. The calculated phonon dispersions (Fig. 6) demonstrate its dynamic stability, at least up to 30 GPa, as there are no imaginary phonon frequencies in this pressure range. As far as we know, there are no other calculations of lattice dynamics of B20-RhGe except ours. No experimental studies are available either. The general shape of the phonon spectrum is very similar to the one calculated [25, 26] and measured [26] for B20 TM monosilicides. The pressure evolution of the phonon dispersions and phonon density of states is shown in Fig. 6. As is seen in the figure, a normal-pressure gap between the optical low- and high-frequency modes disappears on compression.

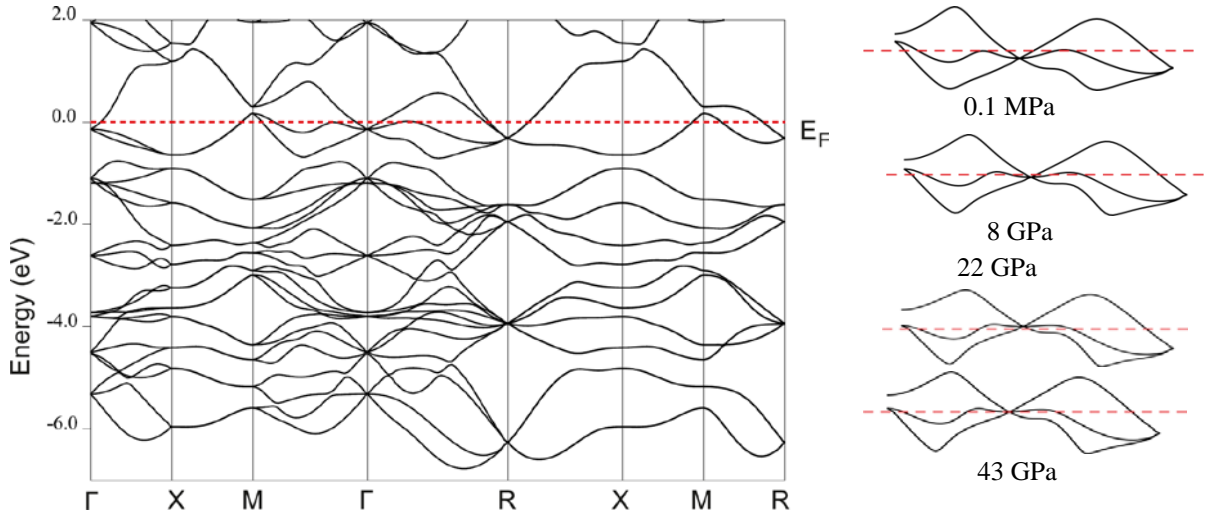

**Fig. 4.** Left: Electronic band structure of B20-RhGe along high-symmetry directions at normal pressure ($P = 0.1$ MPa); the Fermi energy, $E_F$, is set to zero and shown by the horizontal dashed line. Right: Zoom-in of the region of Dirac-point-like feature near $E_F$ at compressions $V/V_0 = 1.0$, 0.95, 0.9, and 0.85 ($V_0$ is the equilibrium volume), the corresponding pressure values are indicated in the figure.

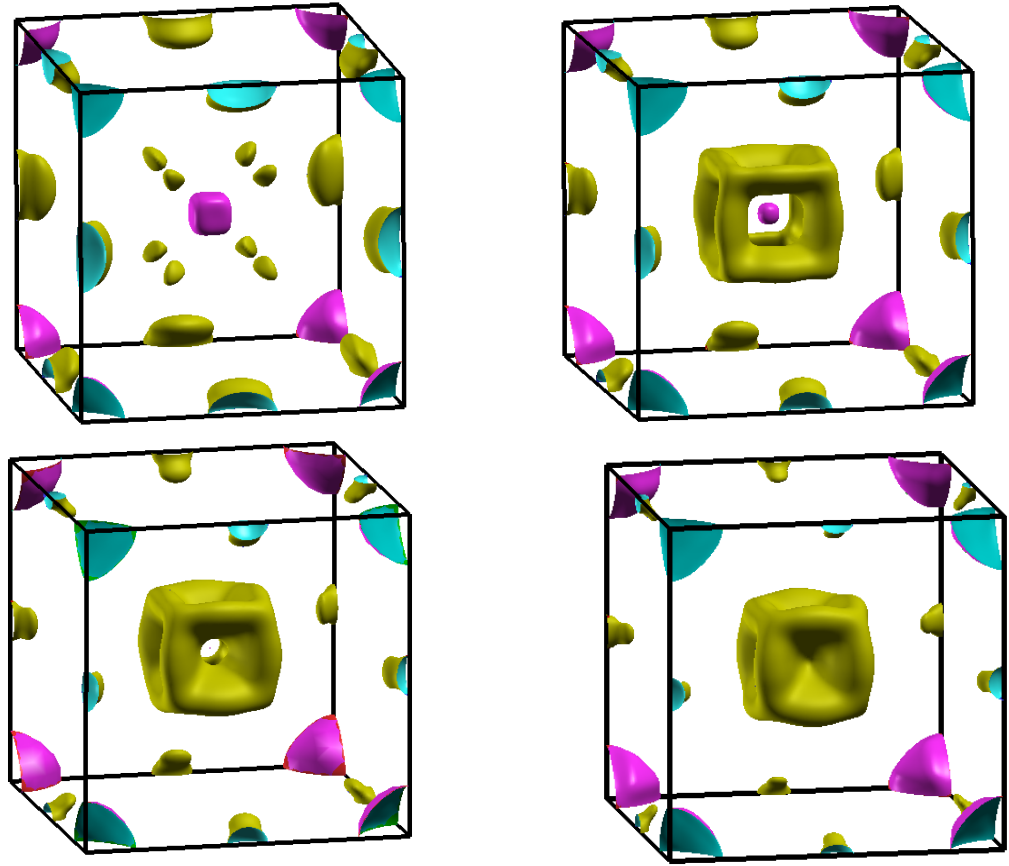

**Fig. 5.** Fermi surface of B20-RhGe at the pressures of 0.1 MPa, 8 GPa, 22 GPa, and 43 GPa. The small electron pocket on the Γ point associated with the Dirac-point-like feature of the electronic spectrum decreases under pressure and finally disappears. The Γ-centered cage-like sheet of holes appears at 8 GPa and then develops into the closed simply-connected surface.

Noteworthy is the pressure behavior of the transverse acoustic modes T1 and T2 propagating along the <110> (ΓM) direction. These modes approach each other upon compression and practically merge at 30 GPa (Fig. 6). This fact should also manifest itself in the pressure behavior of elastic constants, which link together the mechanical and dynamic properties of a crystal. Due to a lack of both experimental and theoretical information on the phonon spectra of B20 compounds under pressure, we compare our results to the calculation [27] of elastic constants in isovalent B20-RhSi, which have been evaluated up to $P = 35$ GPa.

In the $<\xi\, \xi\, 0>$ branch, the sound velocities $v_{T1,2}$ can be expressed through the elastic constants $c'$ and $c_{44}$, associated, respectively, with tetragonal and trigonal shear: $v_{T1}^2 = c'/n$ and $v_{T2}^2 = c_{44}/n$, where $n$ is the density. We have observed a qualitative correlation between the pressure behavior of sound velocities $v_{T1,2}$ of RhGe and elastic constants $c'$ and $c_{44}$ of RhSi. As pressure increases from 0.1 MPa to 30 GPa, the anisotropy factor, $A = (c_{44}-c')/c_{44}$, of RhSi decreases from 0.24 to 0.03 [27]. This behavior of phonon dispersions probably indicates approaching of the RhGe structure to the more isotropic ideal B20 lattice. As is shown in the inset to Fig. 2, at 30 GPa, the atomic positions of B20-RhGe are already rather close to those of the ideal B20 lattice.

### 3.4 Transport properties

One of the ways to experimentally identify a Fermi-surface electronic topological transition (ETP) is to measure the conductivity and thermopower, which are expected to exhibit peculiarities at the ETP. For example, experimentally observed anomalies in the electrical resistivity (ρ) and Seebeck coefficient (*S*) of fcc Ca, Sr, and Yb have been shown to reflect a metal–semiconductor transition [28]. Upon compression, both quantities noticeably increase and reach a maximum at a certain pressure value (below 15 GPa), the effect being most pronounced in case of Yb. Measurements of ρ and *S* under high pressure have been successfully conducted [29] using Toroid device analogous to that employed for the high-pressure synthesis.

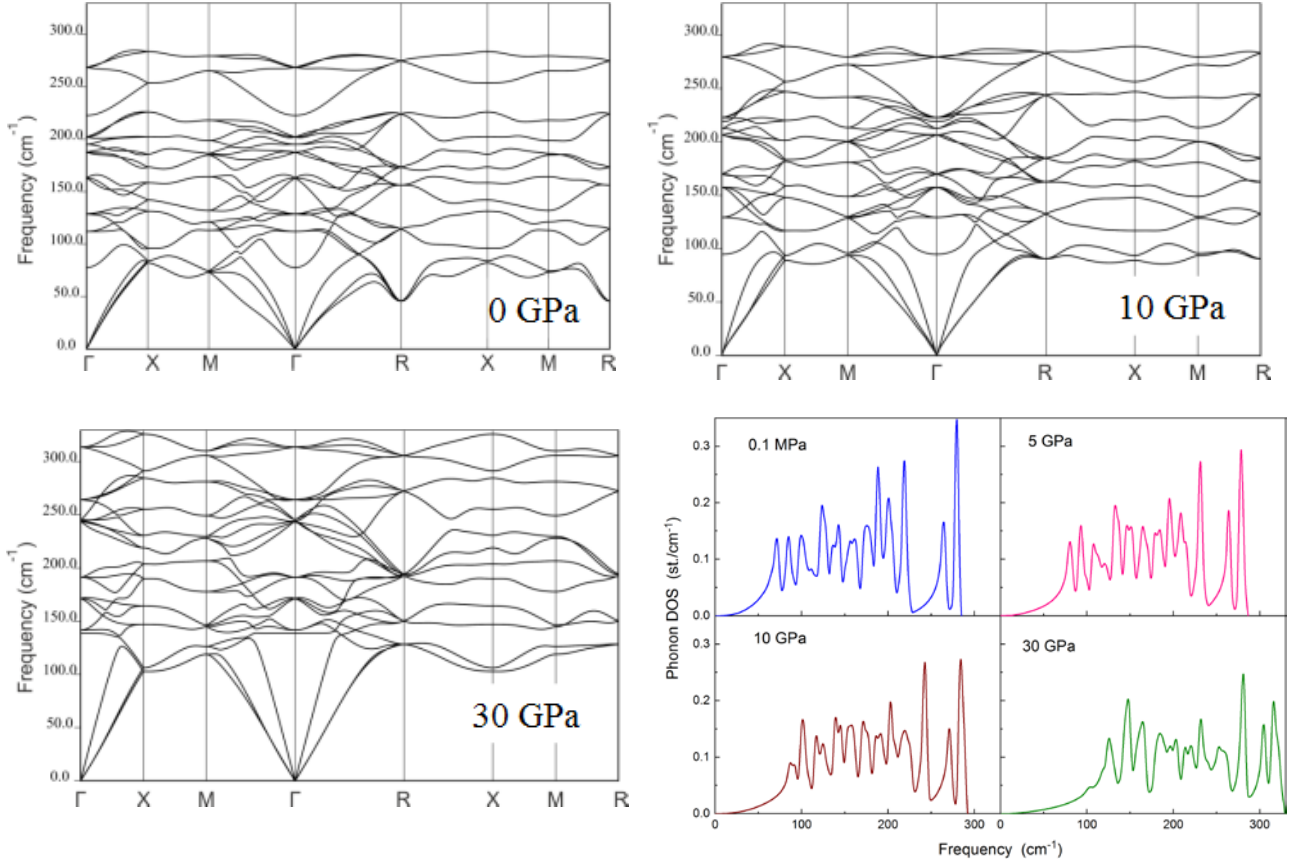

**Fig. 6.** The phonon dispersions and the phonon density of states at ambient and elevated pressures up to 30 GPa.

As is known, the large thermopower *S* is frequently exhibited by materials with asymmetry between *p*- and *n*-type charge carriers near the Fermi energy. Recently, the electrical resistivity and Seebeck coefficient have been theoretically and experimentally studied for B20-type CoGe [9] and CoSi [10, 11] (both isovalent to RhGe), which have been considered as promising thermoelectric materials. A low resistivity and a large negative Seebeck coefficient *S* are found for the both compounds. A negative *S* value indicates that the electron pockets are larger than the hole ones, which leads to the *n*-type carriers dominating the heat transport. The authors [9, 11] studied the possibility of changing the dominating type of carriers by means of partial substitution with electron and hole dopants.

The behavior of electron spectrum similar to the hole-doping effect could be reproduced by applying high pressure. The pressure evolution of FS in B20-RhGe (Fig. 6) suggests the pressure-induced domination of *p*-type carriers, which could be observed experimentally.

The shape of the electronic structure near the Fermi level is very similar for RhGe and for CoGe. This has stimulated us to study the potential of B20-RhGe as a thermoelectric material. Using the BoltzTrap2 code, we evaluate the Seebeck coefficient *S*, the electrical conductivity σ, the electron thermal conductivity κ, and the figure of merit *ZT* of RhGe as functions of temperature and discuss the pressure and doping effect on these properties. In a simple form, the Seebeck coefficient can be expressed by the Mott equation [30]:

$$S = \frac{\pi^2 k_B^2 T}{3e}\left[\frac{d\ln\sigma(E)}{dE}\right]_{E=E_F} = \frac{\pi^2 k_B^2 T}{3e}\left[\frac{d\ln N(E)}{dE} + \frac{d\ln v^2(E)}{dE} + \frac{d\ln\tau(E)}{dE}\right]_{E=E_F}, \quad (1)$$

where *N(E)* is the electron density of states, σ is the electrical conductivity, τ is the relaxation time, and *v* is the average velocity of electrons at the energy *E*. Under a simple assumption that the scattering of electrons is independent of energy, the energy-dependent conductivity is proportional only to the energy derivative of *N(E)*. This formula suggests that a steep slope of the DOS near the Fermi energy is important for the enhancement of *S*. It is just the case of B20 monogermanides, where the hybridization of transition-metal d-electrons with the germanium p- electrons results in a hybridization pseudogap at $E_F$. Due to the Dirac-point-like feature, there is an abrupt drop in the DOS that is large below and small above $E_F$ (see Figs. 3 and 4).

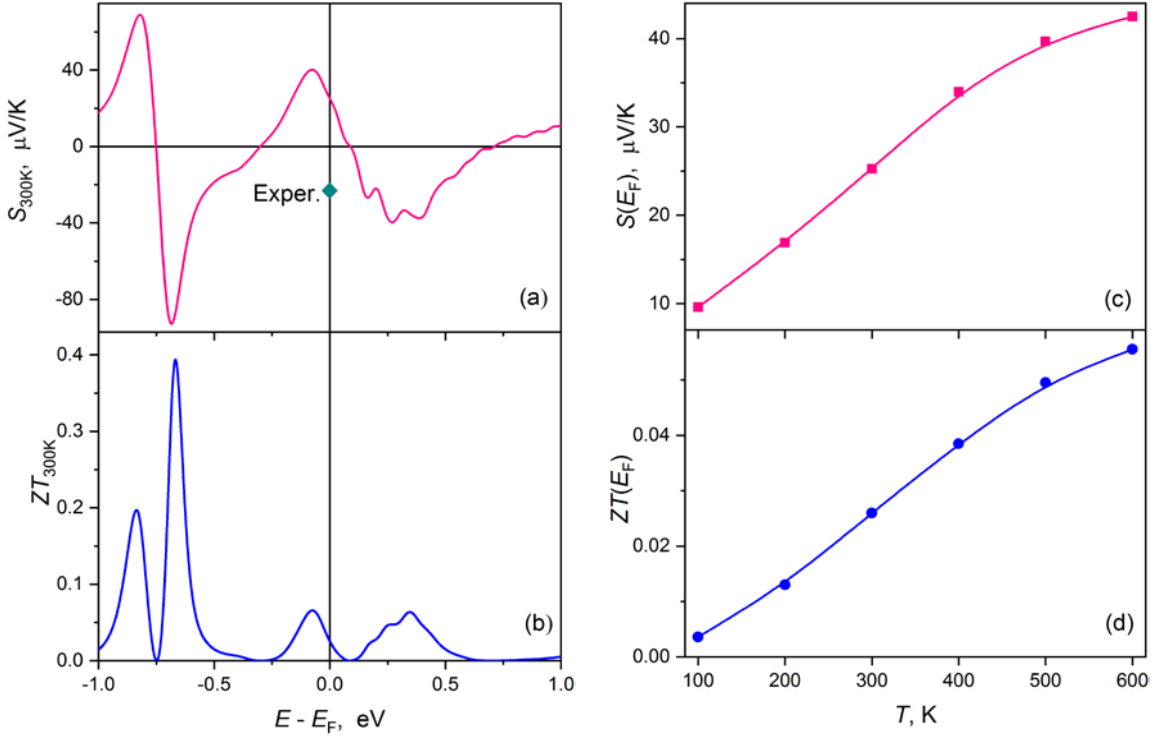

**Fig. 7.** (a, b) The calculated Seebeck coefficient *S* and the figure of merit *ZT* as functions of the position of Fermi energy at $T = 300$ K. (c, d) *S* and *ZT* in dependence of temperature. The experimental room-temperature value of *S* is denoted by the diamond symbol (a).

This also suggests that the thermoelectric efficiency can be improved by tuning the Fermi level to a more suitable position, where the variation in the density of states is strong. The uniform compression shifts the $E_F$ position in the same direction as the Mn, Fe or Ru substitution for Rh (hole doping) does, while the Ni or Pd substitution (electron doping) acts as negative pressure.

In practice, a good thermoelectric material must have high electrical conductivity and Seebeck coefficient and low thermal conductivity κ. Its performance is determined by a dimensionless figure of merit, $ZT = S^2\sigma T/\kappa$. However, the optimization of the thermoelectric performance is not that simple. For example, increasing the Seebeck coefficient *S* usually means decreasing the electrical conductivity and the electron thermal conductivity κ is proportional to the electrical conductivity via the Wiedemann–Franz law.

Figure 7 depicts our calculated Seebeck coefficient *S* and figure of merit *ZT* as functions of Fermi energy and temperature. Experimentally, at room temperature, B20-RhGe exhibits a moderate negative Seebeck coefficient (–23 µV/K) that changes sign at the 2% hole-type doping by manganese (the composition $Rh_{0.98}Mn_{0.02}Ge$) [5]. The theoretical *S* is equal to –23 µV/K at $E = 0.15$ eV and changes sign at 0.09 eV. Thus, the calculated $S(E_F)$ is shifted towards higher energies from the experimental result (a similar shift has been obtained for CoGe [9]).

A reason for this discrepancy is that the DOS around $E_F$ is small forming the pseudogap, and a slight deviation of calculation results may cause a noticeable shift of the $E_F$ position. Another reason is a slightly smaller unit cell volume in the calculations, and this at last might be an indication that an energy dependence of relaxation time should be taken into consideration, as is the phonon part of the thermal conductivity. Anyway, the predicted room-temperature figure of merit is rather small, $ZT = 0.026$. It is also not so large within the pressure (or doping) range of interest, namely, the *S* values accessible on moderate volume compression or doping vary between 40 µV/K and –40 µV/K, while the corresponding *ZT* maximums are reached at 0.35 eV and –0.1 eV and do not exceed 0.07. As the temperature increases, the both *S* and *ZT* rise monotonically, but their values do not change crucially: at 600 K, $S = 42.5$ µV/K and $ZT = 0.056$, which is much lower than values of ~1 expected in good thermoelectrics.

## 4 Conclusions

We performed the comprehensive *ab initio* study of noncentrosymmetric RhGe, a metastable representative of the family of B20-type transition-metal monogermanides. B20-RhGe is synthesized by quenching from the melt under extreme pressures and temperatures of the Earth's mantle and remains metastable at ambient conditions. We evaluated the structural, thermodynamic, electronic, lattice-dynamical, and transport properties of B20-RhGe, including their evolution with pressure and temperature.

The obtained lattice parameters and stability region of B20-RhGe are in agreement with available theoretical and experimental data. A defining feature of the electronic structure is the symmetry-conditioned triply degenerate state (resembling the 3D Dirac point) located just below the Fermi level $E_F$. This feature leads to a steep slope of the density of states, $N(E)$, near the Fermi energy, with a pseudogap typical of poor metals. Our calculations demonstrate that in the rigid band model, a relatively low pressure acts approximately as a low-level hole doping.

The location of the Dirac-point-like feature with regard to the $E_F$ is shown to be essential for the Fermi surface evolution under pressure. As the pressure increases, this feature crosses the Fermi level, which leads to changes in the Fermi surface topology and connectivity (Lifshitz electronic topological transitions). The obtained void- and neck-type Lifshitz transitions are expected to manifest themselves in anomalous pressure behavior of the charge and heat transport properties, which can be observed experimentally.

As is known, the steep slope of $N(E)$ at the Fermi energy (like the one found in RhGe) can lead to an increase in thermopower. With this in mind, using the semi-classical Boltzmann approach in the relaxation-time approximation, we evaluated the electrical conductivity, the electron thermal conductivity, and the Seebeck coefficient ($S$) as functions of temperature and the position of $E_F$, the latter tuned by pressure and doping. Experimental room-temperature Seebeck coefficient has a moderate value of –23 μV/K. The calculated $S(E=E_F)$ is shifted towards higher energies from the experimental result. This discrepancy might be an indication that an energy dependence of the relaxation time should be taken into account, as is the phonon part of thermal conductivity. Anyway, at $T = 300$ K, the predicted thermoelectric figure of merit is rather small, $ZT = 0.026$. Also within the pressure (or doping) range of interest and up to temperatures of 600 K, $ZT$ does not exceed 0.07, which is much lower than expected in good thermoelectrics.

In addition, we have studied the evolution of phonon-related properties with increasing pressure and temperature. The calculated phonon spectra, which still remain to be explored experimentally, demonstrate the dynamic stability of the B20 phase up to at least 30 GPa. At this pressure, the transverse acoustic modes T1 and T2 propagating along the $<\xi\ \xi\ 0>$ direction are found to almost merge. This means a decrease in the elastic anisotropy, that is, the trigonal and tetragonal shear elastic constants, $c_{44}$ and $c'$, approach each other upon compression.

We have constructed the $P$–$T$ phase diagram of RhGe in the quasi-harmonic approximation. The calculated equilibrium line between the high-pressure B20 phase and the low-pressure B31 phase suggests that the former might be synthesized at lower pressures of 4–5 GPa than those currently used. This can reduce the number of structural inhomogeneities and microstresses in the polycrystalline high-pressure phase, which occur during solidification. Hopefully, this approach, being implemented, will enable synthesis of B20-RhGe single crystals.

The support of theoretical calculations by Russian Science Foundation under Grant RSF 18-12-00438 is acknowledged. A.V.T. is grateful to the Russian Foundation for Basic Research for supporting the development of experimental methodology (Grants 17-02-00064 and 17-02-00725). The numerical calculations were carried out using computing resources of the federal collective usage center 'Complex for Simulation and Data Processing for Mega-science Facilities' at NRC 'Kurchatov Institute' (http://ckp.nrcki.ru/) and supercomputers at Joint Supercomputer Center of Russian Academy of Sciences (http://www.jscc.ru). We are also grateful for an access to the URAN cluster (http://parallel.uran.ru) made by the Ural Branch of Russian Academy of Sciences.

## References

[1] L.J. Parker, T. Atou, J.V. Badding, Science **273**, 95 (1996) https://doi.org/10.1126/science.273.5271.95
[2] A.V. Tsvyashchenko, L.N. Fomicheva, M.V. Magnitskaya, V.A. Sidorov, A.V. Kuznetsov, D.V. Eremenko, V.N. Trofimov, JETP Lett. **68**, 908 (1998) https://doi.org/10.1134/1.567954
[3] S.V. Grigoriev, N.M. Potapova, S.-A. Siegfried, V.A. Dyadkin, E.V. Moskvin, V. Dmitriev, D. Menzel, C.D. Dewhurst, D. Chernyshov, R.A. Sadykov, L.N. Fomicheva, A.V. Tsvyashchenko, Phys. Rev. Lett. **110**, 207201 (2013) https://link.aps.org/doi/10.1103/PhysRevLett.110.207201
[4] K. Shibata, X.Z. Yu, T. Hara, D. Morikawa, N. Kanazawa, K. Kimoto, S. Ishiwata, Y. Matsui, Y. Tokura, Nat. Nanotechnol. **8**, 723 (2013) https://doi.org/10.1038/nnano.2013.174
[5] V.A. Sidorov, A.E. Petrova, N.M. Chtchelkatchev, M.V. Magnitskaya, L.N. Fomicheva, D.A. Salamatin, A.V. Nikolaev, I.P. Zibrov, F. Wilhelm, A. Rogalev, A.V. Tsvyashchenko, Phys. Rev. B **98**, 125121 (2018) https://journals.aps.org/prb/abstract/10.1103/PhysRevB.98.125121
[6] J.F. DiTusa, S.B. Zhang, K. Yamaura, Y. Xiong, J.C. Prestigiacomo, B.W. Fulfer, P.W. Adams, M.I. Brickson, D.A. Browne, C. Capan, Z. Fisk, J.Y. Chan, Phys. Rev. B **90**, 144404 (2014) https://doi.org/10.1103/PhysRevB.90.144404

[7] V. Dyadkin, S. Grigoriev, S.V. Ovsyannikov, E. Bykova, L. Dubrovinsky, A. Tsvyashchenko, L.N. Fomicheva, D. Chernyshov, Acta Cryst. B **70**, 676 (2014) https://doi.org/10.1107/S2052520614006611

[8] H. Wilhelm, A.O. Leonov, U.K. Rössler, P. Burger, F. Hardy, C. Meingast, M.E. Gruner, W. Schnelle, M. Schmidt, M. Baenitz, Phys. Rev. B **94**, 144424 (2016) https://link.aps.org/doi/10.1103/PhysRevB.94.144424

[9] N. Kanazawa, Y. Onose, Y. Shiomi, S. Ishiwata, Y. Tokura, Appl. Phys. Lett. **100**, 093902 (2012); http://dx.doi.org/10.1063/1.3691260

[10] A. Sakai, F. Ishii, Y. Onose, Y. Tomioka, S. Yotsuhashi, H. Adachi, N. Nagaosa, Y. Tokura, J. Phys. Soc. Jpn., **76**, 093601 (2007) https://doi.org/10.1143/JPSJ.76.093601

[11] Z.J. Pan, L.T. Zhang, J.S. Wu, Comput. Mater. Sci. **39,** 752 (2007) https://doi.org/10.1016/j.commatsci.2006.09.012

[12] A.V. Tsvyashchenko, V.A. Sidorov, A.E. Petrova, L.N. Fomicheva, I.P. Zibrov, V.E. Dmitrienko, J. Alloys Compd. **686**, 431 (2016) http://dx.doi.org/10.1080/00268979300103121

[13] *Binary Alloy Phase Diagrams,* edited by T.B. Massalski, 2nd edn. (ASM International, Metals Park, Ohio, 1990)

[14] L.G. Khvostantsev, V.N. Slesarev, V.V. Brazhkin, High Press. Res. **24**, 371 (2004) https://doi.org/10.1080/08957950412331298761

[15] I.A. Safiulina, E.V. Altynbaev, E.G. Iashina, A. Heinemann, L.N. Fomicheva, A.V. Tsvyashchenko, S.V. Grigoriev, Phys. Solid State **60**, 751 (2018) https://doi.org/10.1134/S1063783418040273

[16] P. Giannozzi et al., J. Phys. Condens. Matter **21**, 395502 (2009) https://doi.org/10.1088/0953-8984/21/39/395502

[17] J.P. Perdew, A. Ruzsinszky, G.I. Csonka, O.A. Vydrov, G.E. Scuseria, L.A. Constantin, X. Zhou, K. Burke, Phys. Rev. Lett. **100**, 136406 (2008) https://doi.org/10.1103/PhysRevLett.100.136406

[18] A. Togo, I. Tanaka, Scr. Mater. **108**, 1 (2015) http://dx.doi.org/10.1016/j.scriptamat.2015.07.021; A. Togo, L. Chaput, I. Tanaka, G. Hug, Phys. Rev. B **81**, 174301 (2010) https://doi.org/10.1103/PhysRevB.81.174301

[19] G.K.H. Madsen, J. Carrete, M.J. Verstraete, Comput. Phys. Commun. **231**, 140 (2018) https://doi.org/10.1016/j.cpc.2018.05.010

[20] L. Vočadlo, D.G. Price, I.G. Wood, Acta Cryst. **B55**, 484 (1999) https://doi.org/10.1107/S0108768199001214

[21] M. Krajčí and J. Hafner, Phys. Rev. B **87**, 035436 (2013) https://link.aps.org/doi/10.1103/PhysRevB.87.035436

[22] M. Magnitskaya, N. Chtchelkatchev, A. Tsvyashchenko, D. Salamatin, S. Lepeshkin, L. Fomicheva, M. Budzyński, J. Magn. Magn. Mater. **470**, 127 (2019) https://doi.org/10.1016/j.jmmm.2017.10.090

[23] H. Takizawa, T. Sato, T. Endo, M. Shimada, J. Solid State Chem. **73**, 40 (1988) https://doi.org/10.1016/0022-4596(88)90051-5

[24] P. Schlottmann, Eur. Phys. J. B **86**, 101 (2013) https://doi.org/10.1140/epjb/e2013-30550-5

[25] Y.N. Zhao, H.L. Han, Y.Yu, W.H. Xue, T. Gao, EPL **85,** 47005 (2009) https://doi.org/10.1209/0295-5075/85/47005

[26] O. Delaire, K. Marty, M.B. Stone, P.R.C. Kent, M.S. Lucas, D.L. Abernathy, D. Mandrus, B.C. Sales, Proc. Natl. Acad. Sci. USA **108**, 4725 (2011) https://doi.org/10.1073/pnas.1014869108

[27] J.J. Wang, X.Y. Kuang, Y.Y. Jin, C. Lu, X.F. Huang, J. Alloys Compd. **592,** 42 (2014) http://dx.doi.org/10.1016/j.jallcom.2014.01.012

[28] V.E. Fortov, A.M. Molodets, V.I. Postnov, D.V. Shakhrai, K.I. Kagan, E.G. Maksimov, A.V. Ivanov, M.V. Magnitskaya, JETP Lett. **79**, 346 (2004) https://doi.org/10.1134/1.1765180

[29] V.V. Brazhkin, O.B. Tsiok, M.V. Magnitskaya, JETP Lett. **97**, 490 (2013) https://doi.org/10.1134/S0021364013080067

[30] L. Pitaevskii, E. Lifshitz, J. Sykes, *Course of Theoretical Physics: Physical Kinetics* (Elsevier Science, Amsterdam, 2017)